# High-resolution soft-X-ray beamline ADRESS at Swiss Light Source for resonant inelastic X-ray scattering and angle-resolved photoelectron spectroscopies

V.N. Strocov[1,*], T. Schmitt[1], U. Flechsig[1], T. Schmidt[1], A. Imhof[1], Q. Chen[1], J. Raabe[1], R. Betemps[2], D. Zimoch[2], J. Krempasky[2], A. Piazzalunga[1,**], X. Wang[1,3], M. Grioni[3] and L. Patthey[1]

[1] *Swiss Light Source, Paul Scherrer Institute, CH-5232 Villigen-PSI, Switzerland*
[2] *Paul Scherrer Institute, CH-5232 Villigen-PSI, Switzerland*
[3] *Institut de Physique de la Matiére Condensé, Ecole Polytechnique Fédéderale de Lausanne, CH-1015 Lausanne, Switzerland*

**Abstract**

We describe the concepts and technical realization of the high-resolution soft-X-ray beamline ADRESS operating in the energy range from 300 to 1600 eV and intended for Resonant Inelastic X-ray Scattering (RIXS) and Angle-Resolved Photoelectron Spectroscopy (ARPES). The photon source is an undulator of novel fixed-gap design where longitudinal movement of permanent magnetic arrays controls not only the light polarization (including circular and 0-180$^o$ rotatable linear polarizations) but also the energy without changing the gap. The beamline optics is based on the well-established scheme of plane grating monochromator (PGM) operating in collimated light. The ultimate resolving power $E/\Delta E$ is above 33000 at 1 keV photon energy. The choice of blazed *vs* lamellar gratings and optimization of their profile parameters is described. Due to glancing angles on the mirrors as well as optimized groove densities and profiles of the gratings, high photon flux is achieved up to $1\times10^{13}$ photons/s/0.01%BW at 1 keV. Ellipsoidal refocusing optics used for the RIXS endstation demagnifies the vertical spot size down to 4 μm, which allows slitless operation and thus maximal transmission of the high-resolution RIXS spectrometer delivering $E/\Delta E$ better than 11000 at 1 keV photon energy. Apart from the beamline optics, we give an overview of the control system, describe diagnostics and software tools, and discuss strategies used for the optical alignment. An introduction to the concepts and instrumental realization of the ARPES and RIXS endstations is given.

# Introduction

Spectroscopic experiments based on interaction of light with condensed matter belong to the most important tools in contemporary solid state physics. Light in the spectral regions of vacuum ultraviolet and soft X-rays is particularly sensitive to the electronic properties due to its strong coupling to valence electrons in the solid. The ADRESS (ADvanced RESonant Spectroscopy) beamline is a high-resolution undulator beamline constructed for Resonant Inelastic X-ray Scattering (RIXS) and Angle-Resolved Photoelectron Spectroscopy (ARPES) experiments in the soft-X-ray region. The beamline is installed at Swiss Light Source (SLS) on the site of Paul Scherrer Institut, Villigen, Switzerland. SLS is a third generation synchrotron radiation source with an extremely stable storage ring operating in the top-up mode with an electron energy of 2.4 GeV and current of 400 mA. The ADRESS beamline occupies one of 9 straight sections of SLS available for insertion devices.

The ADRESS beamline delivers soft-X-ray radiation with the following parameters:
- photon energy range 300 to 1600 eV;
- circular and linear polarizations, with the linear polarization vector variable within 0-180°;
- resolving power $E/\Delta E$ above 33000 at 1 keV photon energy;
- photon flux at the sample (depending on resolution) between $3 \times 10^{11}$ to $1 \times 10^{13}$ photons/s/0.01%BW at 1 keV photon energy;
- vertical × horizontal FWHM spot size on the sample $3.9 \times 52$ $\mu m^2$ for the RIXS endstation and $10 \times 73.6$ $\mu m^2$ for the ARPES one.

The scientific profile of the beamline is focused on strongly correlated electron systems such as transition metal and rare earth oxides. These systems are nowadays at the forefront of solid state physics due to a variety of their fascinating and practically important properties such as high-temperature superconductivity, metal-insulator transitions, colossal magnetoresistance, etc. High-resolution RIXS and ARPES available at the ADRESS beamline are among the most powerful experimental techniques to study the strongly correlated systems. ARPES experiments in the soft-X-ray energy range benefit from enhanced photoelectron escape depth and thus bulk sensitivity, free-electron final states and better definition of the surface-perpendicular momentum important for 3-dimensional materials. The RIXS endstation includes a high-resolution spectrometer delivering $E/\Delta E$ above 11000 at 1 keV photon energy, which takes the RIXS experiment from the energy scale of charge-transfer and crystal-field excitations to that of orbital and magnetic excitations, and a rotating platform to to study the dispersion of these excitations in **k**-space.

Below we describe in detail the concepts, technical realization and performance of the ADRESS beamline, focusing mostly on the beamline optics.

## Undulator source

The photon source of the ADRESS beamline is the undulator UE44 (for the 44 mm period length) installed in the medium-length straight section 3M of SLS. It has 75 periods, making a total length of 3.4 m. A preliminary account of this device is given in Ref. [1] and a detailed description will be published elsewhere. Briefly, UE44 is based on the APPLE II design with permanent magnets [2]. All four arrays can be shifted in the longitudinal direction, which delivers full control of the light polarization (including circular and linear with the polarization vector variable within 0-180$^o$). Furthermore, following the original idea in Refs. [3], the longitudinal shifts of the four arrays can also control the energy without changing the undulator gap. Our UE44 is the first practical realization of this fixed-gap concept.

Fig. 1 shows calculated flux within the coherent cone of radiation for various polarizations. The flux curves for all linear polarizations are identical, but they start from different energies. The source formed by UE44 is characterized by vertical × horizontal FWHM spot size of $s_V \times s_H = 33 \times 252$ μm$^2$ and angular divergence of the coherent cone of $s'_V \times s'_H = 33 \times 111$ μrad$^2$ around 1 keV photon energy. Increase of the diffraction contribution towards lower energies affects mostly $s'_V$ which increases to 59 μrad at 300 eV.

## Beamline optics

**1. Optical scheme**

The beamline optics is based on the proven scheme of plane grating monochromator (PGM) operating in collimated light [4]. The PGMs in general [5,6] benefit from high resolution, high flux due to absence of the entrance slit allowing the beamline to accept the whole coherent core of undulator radiation, and wide energy range covered with one grating. The collimated light operation of PGM, compared to the original SX700 scheme [7] and later VLS grating schemes [8], brings an additional flexibility [4]: One can steer the PGM through different operation modes characterized by high resolution, high flux or maximal high-order suppression by tuning the cosine ratio

$C_{\text{ff}} = \dfrac{\cos\beta}{\cos\alpha}$ between the incidence and exit angles at the grating α and β. On the practical side, our choice of the optical scheme was supported by good experience with the collimated light PGMs at BESSY as well as with two other beamlines of this type operating at SLS [9].

The optical layout of the ADRESS beamline is shown in Fig. 2. The source is the UE44 undulator. The front end baffles select the coherent cone of radiation. The collimating mirror (CM) is a toroid which converts the divergent beam from the undulator source into a beam which is parallel in the vertical (dispersive) plane. The monochromator situated downstream consists of a plane pre-mirror (PM) and selectable gratings, dispersing the beam in photon energies. Three gratings with constant groove densities *N* of 800, 2000 and 4200 l/mm are used to provide even coverage of the beamline resolution and transmission parameters (see below). Downstream the monochromator is the focussing mirror (FM). Due to radiation safety the optical elements up to FM are enclosed in a hutch with led walls. FM is a cylinder which focuses the dispersed collimated beam on the exit slit, producing monochromatic light. In the horizontal (non-dispersive) plane, the FM has no focussing properties and the beam from the undulator is directly focused by the CM on the exit slit, producing a stigmatic focus. The first refocusing mirror (RM1) is a toroid which refocuses the beam on the sample in the first (ARPES) endstation. When the RM1 is taken away from the beam path, the second refocusing mirror (RM2) having an ellipsoidal shape refocuses the beam on the second (RIXS) endstation. Parameters of all optical elements are summarized in Table 1.

It should be noted that an alternative horizontal focusing scheme is possible where the CM is a cylinder and the focusing is performed by a toroidal FM. Due to larger demagnification this scheme allows reduction the horizontal spot size. However, this is associated with an increase of the horizontal beam divergence at the exit slit, in our case ~40%. The corresponding increase of the light footprints at the refocusing optics is prohibitive for the ellipsoidal RM2 because of technological difficulties to manufacture large high-quality ellipsoidal surfaces. Moreover, this scheme slightly worsens the resolution (see below). Other horizontal focusing schemes involving intermediate collimation between toroidal CM and FM similarly to the vertical focusing [9] are less practical because of additional coupling between the two mirrors which complicates the beamline alignment.

## 2. Resolution optimization

The parameters and actual positions of the optical elements in Fig.2 were optimized in pursuit of high resolving power $E/\Delta E$ above 30000 at 1 keV photon energy. Ray-tracing calculations were performed with the code PHASE [10]. The source was taken to have vertical x horizontal FWHM size $33 \times 252$ μm$^2$ and divergence $33 \times 111$ μrad$^2$ which correspond to the electron beam limit (see above; the diffraction limit had an insignificant effect on the beamline resolution even at low energies). The slope errors (SEs) of the optical elements were the measured values from Table 1.

The beamline resolution $\Delta E$ has the following contributions: (1) The spot size limited resolution $\Delta E_{spot}$ determined by imaging of the photon source onto the exit slit. It is found from the ray-tracing calculations as $\Delta E_{spot} = S_V \cdot \frac{E^2}{1.24} \cdot \frac{\cos\beta}{k \cdot N \cdot f}$, where $S_V$ is the calculated vertical FWHM spot size in μm, photon energy $E$ in eV, $\beta$ the exit angle at the grating, $k = 1$ the diffraction order, and $f$ the exit arm (the formula adopted from Ref. [11]); (2) Diffraction limited resolution $\Delta E_{diff}$ due to the diffraction limit on the grating. Normally minor compared to $\Delta E_{spot}$, this contribution is determined by the expression $\frac{E}{\Delta E_{diff}} = N \cdot l$ [12] which represents the number of grooves within the length $l$ illuminated by the coherent cone of the undulator radiation (restricted by the optical surface); (3) Real operation of the beamline with non-zero slit introduces another contribution, the exit slit limited resolution $\Delta E_{slit}$, which is determined by a formula identical to that for $\Delta E_{spot}$ with $S_V$ replaced by the slit width $d$ [11]. The total resolution $\Delta E$ is the geometrical sum of the three contributions $\Delta E = \sqrt{\Delta E_{spot}^2 + \Delta E_{diff}^2 + \Delta E_{slit}^2}$. Conventionally, our further resolution analysis takes into account only the $\Delta E_{spot}$ and $\Delta E_{diff}$, reflecting therefore the ultimate beamline resolution in the limit of zero exit slit opening.

Our ray-tracing analysis has suggested the following measures to reach the requested resolution:

• For the collimated-light PGMs the resolution is limited mainly by SEs of optical elements rather than aberrations. To identify the most critical ones, we first performed reference ray tracing calculations for an energy of 1 keV and $C_{ff} = 10$ with typical manufacturing SE values, which are around 0.1/0.5 arcsec in the merigional/sagittal directions for the plane optics and 0.5/2.5 arcsec for the toroidal optics. These calculations returned $E/\Delta E \sim 20100$. Then we checked the increments of $E/\Delta E$ upon setting each SE to zero. As already known for the PGMs, the most critical was the meridional SE of the grating, which showed a jump of $E/\Delta E$ by ~26400 relative to the reference value. This SE had to be specified below 0.07 arcsec, close to the present technological limit, in order to reach the requested beamline resolution. The second critical was the sagittal SE of the FM, where the increase was ~1700. It was specified to be below a more relaxed value of 1.5 arcsec. Other SEs had insignificant effect on the resolution;

• The focus on the exit slit was made stigmatic in order to minimize influence of the grating and exit slit misalignments on resolution;

• The horizontal focussing is performed by the toroidal CM through the cylindrical FM. Our ray-tracing analysis has shown that this improves maximal $E/\Delta E$ by ~1000 compared to the alternative scheme, where the horizontal focusing performed by a toroidal FM (see above).

### 3. Resolution parameters

Results of the resolution calculations with the optimized beamline geometry are shown in Fig. 3. First, the insert displays a typical spot at the exit slit generated by ray-tracing calculations with the grating 4200 l/mm and $C_{ff} = 4$. The vertical × horizontal FWHM spot size is $S_V \times S_H = 14.1 \times 228$ µm². Its theoretical limit is set by direct demagnification of the undulator source by the ideal beamline with zero SEs and aberrations. With our optical scheme, the demagnification in the vertical directions is given by $\frac{r'}{rC_{ff}}$, where $r$ and $r'$ is the source-to-CM and FM-to-slit distances respectively, and in the horizontal direction by $\frac{r''}{r}$ where $r''$ is the CM-to-slit distance. This sets the theoretical limit at $S_V \times S_H = 4.4 \times 185$ µm². The additional broadening for the real beamline is due to mostly the SEs (predominantly the meridional SE of the grating, see above) rather than aberrations, which is reflected by the Gaussian profile of the spot.

The plots in Fig. 3 show the calculated $\Delta E$ (including the $\Delta E_{spot}$ and $\Delta E_{diff}$ contributions) represented as $E/\Delta E$ depending on photon energy. The three bunches of lines correspond to the three gratings, and variation of $E/\Delta E$ through each bunch to variation of $C_{ff}$ by integer values between the indicated limits. The low limit is chosen here as the floor function of the flux-optimal $C_{ff}$ (delivering maximal overall flux in the region 700-1200 eV, see below) and the high limit as the maximal integer $C_{ff}$ for which the light footprint on the grating does not split beyond its optical surface of 90 mm through the whole shown energy range. The increase of $C_{ff}$ results in improvement of resolution due to increasing demagnification of the photon source. The plot shows that for the 4200 l/mm grating and $C_{ff}$ = 10 the theoretical $E/\Delta E$ is about 33700 at 1 keV photon energy. The shaded bands in the plot display $E/\Delta E$ determined by $\Delta E_{spot}$ without the diffraction contribution $\Delta E_{diff}$. The edges of the bands correspond to the $C_{ff}$ limits of the above bunches. Obviously, the diffraction contribution becomes significant at low energies and small $C_{ff}$ values when the incidence angle on the grating becomes less grazing.

The theoretical resolution was verified by characteristic X-ray absorption spectra of a few gases. The measurements were performed in a gas cell with gas pressure around $1 \times 10^{-2}$ mbar. The monochromator was set to $C_{ff}$ = 2.15 and exit slit to 10 μm. In Fig. 4,

- (*a*) shows a typical absorption spectrum of $N_2$ at the $1s \rightarrow \pi^*$ resonance. Its fine structure is due to vibrational levels of the $\pi^*$ state. The spectrum was measured with the grating 800 l/mm. We have estimated $\Delta E$ (now including the $\Delta E_{slit}$ contribution) by comparing the experimental ratio of the first minimum to the third maximum [13] to that in a simulated spectrum [14]. A Lorentzian width of 112 meV was assumed, which is the lower limit of the diverging data found in literature. Such a conservative estimate demonstrated $E/\Delta E$ at least better than 8500 which is consistent with the theoretical $E/\Delta E$ about 9600 for these beamline settings (including $\Delta E_{slit}$ ~ 13 meV). This figure is close to the limit of resolution sensitivity of the $N_2$ spectra, which is limited by some uncertainty of the Lorentzian width as well as by noise in the experimental spectra. We could hardly identify any resolution improvement with higher-$N$ gratings.

- (*b*) shows a spectrum of Ne at the 1s Rydberg series measured with the same grating 800 l/mm. To estimate the resolution, we fitted the first spectral peak with a Voigt profile having a Lorentzian width of 252 meV as reported in Ref. [15]. This yielded $E/\Delta E$ ~ 8000 which again well conforms to the theoretical $E/\Delta E$ about 8400 (including $\Delta E_{slit}$ ~ 43 meV).

- (*c*) shows an O 1s absorption spectrum of CO. Its first peak is the O $1s \rightarrow 2\pi$ resonance with fine structure due to the vibrational levels in the $2\pi$ state. The spectrum was measured with the grating 4200 l/mm. To our knowledge, this is the best resolved spectrum of CO published to date. Fitting the spectrum with a sequence of Voigt profiles with a Lorentzian width of 156 meV [16] yielded $E/\Delta E$ at least better than 15000. This figure is at the resolution sensitivity of the CO spectrum, whereas the theoretical $E/\Delta E$ for these beamline settings is about 36400. With the ARPES endstation coming in operation we will be able to reliably control such ultimate beamline resolutions by measurement of photoemission spectra at the Fermi edge.

Practically, we have found that the beamline resolution is severely affected by vibrations coming with turbulent flow of cooling water through the gratings and especially the PM. Cooling of the sagittaly mounted CM has only negligible effect on the resolution, but causes notable vibrations of the horizontal beam position. It is therefore crucial to keep the water flow through all these optical elements at the minimal necessary level. Apart from the usual manual adjustment for each element, we remotely control the main pump of the cooling system to adjust the total water flow according to the *K*-value of the undulator and thus total heat load on the beamline.

**4. Transmission optimization**

As both RIXS and soft-X-ray ARPES experiments are characterized by extremely low signal, flux performance of the beamline is of paramount importance. To achieve the best beamline transmission, a few measures have been taken: (1) Most glancing incidence angles at all mirrors (see Table 1); (2) Grating were chosen with minimal *N* necessary to achieve the requested resolutions balanced with the endstation resolutions; (3) Optimization of the grating profile parameters.

Profiles of the gratings were optimized using calculations of their reflectivity with the code REFLEC [17]. First, we had to determine whether the gratings should be lamellar or blazed. We compared the two options for the gratings 800 and 2000 l/mm. The calculations were performed for the ideal grating profiles with the apex angle $90^o$, and realistic profiles with apex angles around $170^o$ for the blazed gratings and $165^o$ for the lamellar ones (data by ZEISS). The profile parameters were optimized for an energy of 930 eV and $C_{ff}$ of 2.25: the blaze angle $\phi_{blaze}$ was $0.8^o$ for the 800 l/mm grating and $1.3^o$ for the 2000 l/mm one; for the lamellar option the groove depth *h* and duty cycle *c/d* (at the groove half-height for the realistic profiles) were 11.0 nm and 0.69 correspondingly for the 800 l/mm grating, and 5.5 nm and 0.6 for the 2000 l/mm one (the procedure to determine *h* and *c/d* is described below). An *rms* surface roughness of 0.5 nm was assumed in all cases.

Fig. 5 shows the calculated energy dependences of reflectivity *R(E)*. The blazed gratings demonstrate remarkable flatness of *R(E)* which is, surprisingly, even superior to their lamellar counterparts. For the 800 l/mm case, the blazed profile increases *R(E)* compared to the lamellar through the whole energy region from 300 to 1800 eV. In the case of ideal grating profiles (*solid*

*lines*) the increase is on average about a factor of 2. Introduction of the realistic profiles (*dashed lines*) does not much change *R(E)* of both gratings, and the blazed stays in immense advantage. Due to this analysis for the 800 l/mm grating we have opted for the blazed profile with an idea to deliver maximal flux at moderate resolution. The fact that the blazed gratings suffer from worse higher order suppression (HIOS) compared to the lamellar, was less important for our beamline operating at rather high energies. For the 2000 l/mm grating, Fig. 5 shows that in the ideal case the increase of *R(E)* due to the blazed profile is again about the factor of 2. However, when the realistic profiles are introduced, the blazed profile shows much stronger degradation of *R(E)* (physically, the degradation occurs due to decrease of the reflecting facets area with increase of the apex angle and *N*) compared to its lamellar counterpart, reducing the advantage to a factor about 1.5. We considered this less significant compared to more complicated manufacturing process and thus higher prices and potential quality problems of the blazed gratings. For the 2000 l/mm and, similarly, 4200 l/mm gratings we have therefore opted for the lamellar profile.

The second step was to optimize the *h* and *c/d* parameters of the lamellar gratings. Commonly the optimization is performed by the manufacturer for a requested energy and $C_{ff}$ value. However, the latter is a priori unknown, because its flux optimal value significantly depends on energy, *N* and profile parameters of the grating (the traditional $C_{ff}$ = 2.25 goes back to the original SX-700 monochromator [7] where this value was optimized for a particular energy range around 700 eV and grating of 1200 l/mm). The optimal $C_{ff}$ should therefore be determined in an optimization procedure simultaneously with *h* and *c/d*. Furthermore, the variation of $C_{ff}$ implies variation of the deviation angle at the grating and thus at the PM, with consequent variation of reflectivity at this optical element. Therefore, the target function in the optimization procedure should be the total transmission $R_2 = R_{PM} * R_{grating}$ of the PM and grating pair.

The optimization used the transmission $<R_2>$ averaged over 8 points in an energy interval from 700 to 1200 eV. The apex angle was 164° for *N* = 2000 l/mm and 163° for *N* = 4200 l/mm (data by ZEISS) and *rms* roughnes was 0.5 nm. Fig. 6 (*top*) shows $<R_2>$ calculated for the 2000 l/mm grating over a three-dimensional grid of *h*, *c/d* (in the following *c* refers to the top of the grooves, which is standard for the input of REFLEC) and $C_{ff}$. The three crossing planes identify the broad transmission maximum centered at *h* = 5.7 nm, *c/d* = 0.645 and $C_{ff}$ = 3.125. Fig. 6 (*bottom*) shows an additional optimization criterion, flatness of the $<R_2>$ energy dependence characterized by relative *rms* deviation $<\Delta R_2>/<R_2>$ calculated over the above 8 points. Obviously, it gradually improves with decrease of *h* and *c/d* and reaches its minimum at their smallest values. We have therefore chosen the optimal *h* and *c/d* values slightly smaller than in the $R_2$ maximum, which notably improved the flatness at insignificant (<0.5%) degradation of $R_2$. However, the minimum of $<\Delta R_2>/<R_2>$ appears at the nearly the same $C_{ff}$ value as the $<R_2>$ maximum. This analysis yielded the flux-optimal parameters for the 2000 l/mm grating as $h = 5.5$ nm, $c/d = 0.63$ and $C_{ff} = 3.125$. The same optimization for the 4000 l/mm grating has yielded the $R_2$ maximum at $h = 4.75$ nm, $c/d = 0.65$ and $C_{ff} = 4.75$. To improve the flatness, the flux-optimal parameters for this grating were chosen as $h = 4.5$ nm, $c/d = 0.64$ and $C_{ff} = 4.75$. As mentioned above, optimization of the gratings on HIOS was not a concern for our beamline.

We have also determined the flux-optimal $C_{ff}$ for the 800 l/mm blazed grating by optimization of $<R_2>$ on a one-dimensional grid of $C_{ff}$ with the blaze angle varying simultaneously with $C_{ff}$ according to $\varphi_{blaze} = \dfrac{\alpha + \beta}{2}$. The maximum was found at $C_{ff} = 2.15$ and $\varphi_{blaze} = 0.8°$.

**5. Flux parameters**

Fig. 7 shows the theoretical flux the beamline delivers at the sample (after the RM) as a function of photon energy for LH polarized light from the undulator with a crossover between the 1st and 3rd harmonics around 865 eV (Fig. 1). Similarly to the resolution data in Fig. 3, the three bunches of solid lines correspond to the three gratings. Variation of flux from the bottom to top edges of each bunch corresponds to variation of $C_{ff}$ in integer values the same as in Fig. 3 (from the floor function of the flux-optimal $C_{ff}$ in the 700-1200 eV region to the maximal integer $C_{ff}$ for which the light footprint stays within the optical surface of the gratings). The flux is normalized to the energy-dependent bandwidth corresponding to the $E/\Delta E$ values typical of each grating.

Another bunch in Fig. 7 shown shaded represents the same flux curves calculated for the 800 l/mm grating operating in the 2$^{nd}$ diffraction order. Compared to the 2000 l/mm grating in the 1$^{st}$ order, this operation mode, having close resolution parameters, delivers higher flux in the high-energy region above 1 keV. Such a high 2$^{nd}$ order intensity is due to the blazed profile of the 800 l/mm grating.

A closer look on the flux curves in Fig. 7 shows that the flux-optimal $C_{ff}$ value increases with energy and groove density. This is illustrated by calculated energy dependences of the flux-optimal $C_{ff}$ in Fig. 8. The energy variation is particularly large for the 800 l/mm grating in the 2$^{nd}$ order and 4200 l/mm grating.

Fig. 7 shows that at 1 keV photon energy and flux-optimal $C_{ff}$ value the theoretical flux for the 800 l/mm grating is $1.3 \times 10^{13}$ ph/s in a bandwidth corresponding to $E/\Delta E = 10000$, for the 2000 l/mm one $1.7 \times 10^{12}$ ph/s to $E/\Delta E = 15000$, and for the 4200 l/mm one $2.4 \times 10^{11}$ ph/s to $E/\Delta E = 20000$. We have experimentally confirmed these flux figures by measurements with a photodiode (AXUV100 from International Radiation Detectors) using the responsivity curves available from the manufacturer [18]. Compared to one of the best up to date high-resolution soft-X-ray beamlines BL25SU at Spring-8 [19] with the flux values normalized to the same bandwidth, the ADRESS beamline delivers flux higher by about two orders of magnitude for $E/\Delta E = 10000$ and about an order for $E/\Delta E = 15000$. Such an excellent flux performance has resulted from a few factors: the energy of the ring optimized for the soft-X-ray photon energy range, optical scheme of PGM without entrance slit allowing acceptance of the whole coherent core of undulator radiation, and all above measures to optimize the beamline transmission.

**6. Refocusing optics**

The refocusing was particularly demanding for the RIXS endstation because slitless operation of our high-resolution RIXS spectrometer (see below) required to squeeze the vertical spot size below 5 μm FWHM. In order to reduce flux loss, we have restricted ourselves to one-mirror refocusing systems and small grazing incidence angle $\alpha = 1°$. We have investigated two options, a toroidal refocusing mirror (T-RM) and ellipsoidal one (E-RM). It should be noted that the ellipsoidal refocusing optics can only be used if the beamline focus is stigmatic.

Model ray-tracing calculations were performed with the source taken as the typical spot produced by the beamline at the exit slit (see above) which has the dimensions $S_V \times S_H = 14.1 \times 228$ μm$^2$ (such a source is roughly equivalent to the exit slit open to a width of $S_V$). The source to the sample distance was taken as 7000 mm, and the meridional/sagittal FWHM SEs as their typical values 0.5/1.5 arcsec for the T-RM and three times those for the E-RM. The resulting spot size at the sample $w$ has 3 contributions: nominal demagnification of $S_{V,H}$ by the ratio $r/r'$ of the object and image distances as $S_{V,H} \cdot \frac{r}{r'}$, SEs of the mirror $w_{V,H}^{se}$ and aberrations $w_{V,H}^{aberr}$, for both vertical and horizontal directions. The contributions add up geometrically as

$$w_{V,H} = \sqrt{\left(S_{V,H} \cdot \frac{r}{r'}\right)^2 + \left(w_{V,H}^{se}\right)^2 + \left(w_{V,H}^{aberr}\right)^2} \; .$$

Fig. 9 (*top panel*) shows dependences of the vertical spot size $s_V$ on the nominal demagnification $r/r'$ calculated for the T-RM in comparison with the E-RM. In the region of small $r/r'$, the former outperforms the later because of smaller SEs. With increase of $r/r'$, however, $w_V$ produced by the T-RM rapidly reaches its minimum of ~10 μm at $r/r'$ ~ 1.8, where the growing $w_V^{aberr}$ contribution starts to prevail over the decreasing $S_V \cdot \frac{r}{r'}$ one, and the spot starts to blur (see also Ref. [6]). The situation is different for the E-RM, which is in principle the ideal point-to-point focuser: Due to smaller aberrations $w_V$ carries on decrease towards much larger $r/r'$ and reaches its minimum of ~3.4 μm at $r/r'$ ~ 9. The use of the E-RM allows thus more effective demagnification at large $r/r'$.

Fig. 9 (*bottom*) shows the same comparison for the horizontal spot size $w_H$. This case is less critical on aberrations, because due to rather large $S_H$ the dominating contribution in $w_H$ is $S_H \cdot \frac{r}{r'}$ which drives it to decrease towards higher $r/r'$. The minimum of $w_H$ is found at $r/r'$ ~ 4 for the T-RM and well above 10 for the E-RM.

In our beamline the RIXS endstation takes the furthest position from the exit slit, which allows $r/r'$ up to 5.85 (see the optical scheme in Fig. 2). As small as possible $w_V$ should be achieved for the slitless spectrometer operation. We have therefore opted here for refocusing with the E-RM. With the actual SEs of our E-RM (Table 1) and $r/r'$ = 5.85 this delivers a spot size of $w_V \times w_H = 3.9 \times 52$ μm$^2$ for the above typical spot of 14.1 μm at the exit slit, and the minimal spot size of 3.1 μm in the limit of zero exit slit opening. We have fully confirmed such a small spot size by measurements with the RIXS spectrometer (see below) which in the zero-order diffraction mode allows direct imaging of the spot on the CCD camera.

Requirements of the ARPES endstation on the spot size are mainly due to angular resolution of the ARPES spectrometer. In our case the ultimate resolution is achieved with the spot size less than 100 μm in both directions. With $S_H$ being much larger than $S_V$, the horizontal demagnification becomes more critical than the vertical one. Furthermore, the ARPES station is placed the first from the slit, which restricts us to relatively small $r/r'$ up to 3.28. Due to these two considerations a satisfactory performance can be achieved here with the simpler T-RM option, which for $r/r'$ = 3.28 shows still improvement of $s_H$ without much degradation of $s_V$. With the actual SEs of the T-RM (Table 1) the spot size on the sample was calculated as $w_V \times w_H = 10.0 \times 73.6$ μm$^2$. Anyway, based on our good experience with the E-RM for the RIXS endstation and in view of importance of further reduction of $w_H$ for experiments on mosaic samples we plan in future to increase $r/r'$ and upgrade to the ellipsoidal mirror.

On the practical side, it has been crucial that the manufacturer was able to produce E-RM with extremely small SEs (see Table 1). Furthermore, it should be noted that the ellipsoidal optics is by far more sensitive to alignment compared to the toroidal one. We have therefore mounted the refocusing mirrors on advanced hexapod mechanics from Oxford-FMB [20] delivering 3 translational and 3 rotational fully decoupled degrees of freedom (DOFs) with a practical accuracy of 1 μm and 1 μrad respectively.

**7. Diagnostics**

The beamline is equipped by the whole bundle of standard diagnostics tools such as X-ray beam position monitors (XBPMs) in the front end, retractable fluorescent screens along the beam path (downstream the FM as well as in front of both RMs), the AXUV100 photodiode after the exit slit, drain current readouts at the RMs, etc.

Extremely useful for the beamline alignment and monitoring proved to be an originally developed online beam monitor installed in front of the exit slit. The monitor uses a fluorescent YAG crystal with dimensions of 24×24 mm$^2$ installed slightly above the slit. The beam produces here a vertical line of energy dispersed light. This image is streamed in over the network by an IP camera with a resolution of 640×480 pixel (AXIS210 from AXIS Communications) and is visualized with a internet browser. A real-time image processing software written in Matlab (Fig. 10) intercepts the stream, selects the region of interest around the line of light (*left panel*), performs compensation of the line curvature due to optical aberrations, sums up the image rows through the region to evaluate the average horizontal profile, performs Gaussian fit of the profile to evaluate the horizontal beam position and width (*right*) and streams these parameters as process variables into the beamline control system. Although the data processing algorithm includes compensation of the standard gamma-correction embedded in the camera, one should be careful with an intensity analysis because the YAG crystal has a non-linear intensity response. The code is available from the first author.

The above beam monitor is particularly useful with our optical scheme (Fig. 2) where the horizontal focusing at the exit slit is performed by the CM decoupled from the cylinder FM. In this case monitoring of the horizontal beam width allows us to adjust the pitch angle of the CM. Due to coordinated horizontal and vertical curvatures, this ensures that the CM produces vertically collimated light.

**8. Optical alignment strategies**

Alignment of the beamline is a complex process of optimization in a multidimensional space of the translational and angular DOFs of all optical elements. Otherwise hopeless, successful alignment requires that the optimization process is broken into well-defined procedures with reduced dimensionality. As an example, we here present analysis of vertical focusing of the beamline, directly related to the energy resolution, and demonstrate a technique to reduce this problem to fast and straightforward optimization in one dimension.

A simplified scheme of vertical focusing is given in Fig. 11. To focus the beam on the exit slit, one tweaks the FM pitch $R_y^{FM}$. This immediately displaces the horizontal beam position from the slit center, which should be back corrected by the FM horizontal translation $z^{FM}$. The latter, in turn, displaces the FM aperture away from the beam coming from the CM, requiring correction of the CM pitch $R_y^{CM}$. Commonly, one performs focusing by iterative tweaking of all these three motions, which is a time consuming and less controllable procedure of optimization in three dimensions.

In fact, two conditions that (1) the beam stays on the center of the slit, and (2) the beam from the CM hits the center of the FM aperture, link the individual $R_y^{FM}$, $z^{FM}$ and $R_y^{CM}$ motions into one *combined focalization motion* (CFM) acting as one DOF. We have set up the CFM for our beamline using the following method: The $z^{FM}$ motion is chosen as a parameter of the CFM. The remaining two individual motions are taken as linear functions of $z^{FM}$ as

$$R_y^{CM}(z^{FM}) = a_0^{CM} + a_1^{CM} \cdot z^{FM}$$
$$R_y^{FM}(z^{FM}) = a_0^{FM} + a_1^{FM} \cdot z^{FM}$$

To determine the coefficients $a$, we take two arbitrary reference $z^{FM}$ points separated by a few mm. First, in each point we find the $R_y^{CM}$ value corresponding to the center of the FM aperture. For this purpose we scan the aperture with $R_y^{CM}$ and detect the transmitted beam on the YAG crystal in front of the slit using the above beam monitor. (It may be necessary to roughly adjust $R_y^{FM}$ in so that the whole excursion of the beam stays within the YAG crystal; the photodiode behind the slit was not suitable as a detector for this scan, because its size was smaller than the excursion). Having determined the reference $R_y^{CM}$ values for the two $z^{FM}$ points, we immediately find the above coefficients $a_{0,1}^{CM}$ with a linear fit. Second, in each reference $z^{FM}$ point we set the determined $R_y^{CM}$ values and tweak $R_y^{FM}$ to put the beam exactly at the center of the slit. The two reference $R_y^{FM}$ determined in this way immediately fix the coefficients $a_{0,1}^{FM}$. The CFM defined with this method was realized using a simple GUI-based software which upon input of the $z^{FM}$ parameter automatically calculates the corresponding $R_y^{FM}$ and $R_y^{CM}$, and drives the motors of all three individual motions. Accuracy of the CFM setup can be judged by constancy of the beam position at the slit when scanning the CFM. Normally, the deviation is less than 75 μm for the whole focalization span of $z^{FM}$.

The concept of CFM allows reduction of the beamline focalization problem to fast and unambiguous procedure of optimization in one dimension, as compared to the common iterative tweaking in three dimensions. In addition, by the very definition of the CFM, the beam always stays in the middle of the FM optical surface, which ensures the highest resolution due to minimal SEs as well as the highest beamline transmission.

Fig. 12 shows typical measurements of a focalization curve as a resolution factor $F_R$ depending on the CFM parametrized by $z^{FM}$. Here, $F_R$ was quantified from the O 1s absorption spectrum of CO as the averaged minimum-to-maximum ratio of three oscillations on top of the spectrum (window in the insert of Fig. 12). Cubic polynomial fit (*gray line*) of the measured points puts the focal point at $z^{FM} = 1.64$ mm. Typically, we perform full beamline focalization in 1-2 hours.

## Control system

A scheme of the beamline control system is shown in Fig. 13. It consists of two main subsystems:

• Equipment Protection System (EPS) which surveys the temperatures and pressures interlocked to the beamline shutter, valves and water cooling. This subsystem is based on Siemens S7 programmable logic controllers (PLCs).

• Beamline control based on Experimental Physics and Industrial Control System (EPICS) which is a set of open source software tools to create distributed real-time control systems for large-scale scientific instruments such as particle accelerators [21]. On the upper level of our EPICS subsystem are client programs, which run on Linux or Windows workstations. On the lowest level are server programs. Via the network they communicate, on one side, with the clients using so-called channel access protocol provided by EPICS and, on the other side, with real-time computer hardware which is based on VERSAmodule Eurocard (VME) bus (IEEE 1014) and runs the Input-Output Controller (IOC) software. The VME hardware directly controls the motions of all beamline components (the insertion device, monochromator, mirrors, etc.) as well as analog/digital inputs and outputs. The use of the VME hardware allows real-time beamline control independent of the network latency. The two hexapod systems use a different type of real-time hardware, which is based on PMAC architecture running under Linux MicroIOC [22]. The beamline motion control uses stepping motors, which are driven by units operating with the OMS58 Intelligent Motor Controller, and linear encoders from Renishaw Inc.

The EPS subsystem is incorporated into EPICS via mapping the PLCs to IOC channels [23]. An important aspect of our control system is that these channels appear in EPICS on equal footing with other IOC channels which gives a consistent look-and-feel to the overall beamline instrumentation control.

### Endstations

Here we briefly summarize the scientific concepts and technical implementation of our ARPES and RIXS endstations. Their detailed description will be presented elsewere.

ARPES is a photon-in / electron-out technique which probes the electronic structure of solids in the sub-surface region. The ARPES spectra directly characterize the hole spectral function $A(E,\mathbf{k})$ with resolution in energy and 3-dimensional $\mathbf{k}$. In the soft-X-ray range, increased photoelectron escape depth results in enhanced bulk sensitivity, and free-electron final states and reduced broadening in the surface-perpendicular momentum allow reliable studies of 3-dimensional systems. Excellent flux parameters of the ADRESS beamline help the notorious problem of small valence band photoexcitation crossection in the soft-X-ray range.

The ARPES endstation is available for the users from autumn 2009. It features the following main instrumental concepts:

• Light is incident on the sample at a grazing angle of $20^o$, giving a gain in photoelectron yield about a factor of 2 compared to the conventional $45^o$. To balance the horizontal and vertical size of the light footprint, the sample is rotated around the horizontal axis;

• The manipulator (CARVING design by PSI and Amsterdam University delivering 3 mechanically decoupled angular DOFs) is mounted in horizontal orientation of the primary rotation axis. It is equipped by He flow cryostat which allow working temperatures down to 10K;

• The analyser (PHOIBIOS 150 from SPECS delivering an angular resolution about $0.07^o$ and ultimate energy resolution better than 5 meV) is rotatable around the lens axis to change the slit orientation relative to the incoming light: When the slit is in the scattering plane, one can explore symmetry of the valence states using the selection rules with variable polarization; when perpendicular, one can efficiently sample **k**-space by the primary manipulator rotation.

RIXS is a photon-in / photon-out bulk-sensitive technique which delivers information about electronic structure of solids, liquids and gases resolved in atomic and orbital character [24]. The photon energy loss spectra measured in RIXS reflect the spectrum of charge neutral (e.g., crystal field, charge transfer and spin) electronic excitations, and their photon momentum transfer dependence characterizes the dispersion of these excitations in **k**-space. This technique is particularly useful for strongly correlated systems such as transition metal oxides, rare earth systems as well as diluted systems like buried layers or nanostructures.

Our RIXS endstation is in operation since spring 2007. It has the following features:

• A high-resolution spectrometer nicknamed Super Advanced X-ray Emission Spectrometer (SAXES) based on VLS spherical grating. It delivers resolving power better than 10000 in the energy range up to 1100 eV, which for the first time takes the RIXS technique from the energy scale of charge transfer and crystal field excitations to that of orbital and magnetic excitations. The experiments are normally carried out at sample temperatures down to 12 K in order to reduce phonon broadening of the spectral structures. Design and realization of the SAXES spectrometer have been described in detail in Ref. [25]. A method to determine the spectrometer settings to cancel coma aberrations for any energy is described in Ref. [26];

• A platform with 5 DOFs which serves as an optical bench for the spectrometer. The platform is rotatable to vary the scattering angle in 6 steps between $30^o$ and $130^o$. This feature allows variation of photon momentum transfer and thus studies of dispersion of low-energy excitations in **k**-space.

First application examples illustrating the power of high-resolution soft-X-ray RIXS in investigation of orbital and magnetic excitations can be found in Refs. [27,28].

## Summary

We have presented in detail the concepts and technical realization of the high-resolution beamline ADRESS operating in the soft-X-ray energy range from 300 to 1600 eV. Below we summarize the main features of the beamline:

• The undulator source is the first realization of the fixed-gap concept where both polarization and energy of the light are changed by longitudinal movement of permanent magnetic arrays without changing the gap. The undulator gives full polarization control, delivering left- and right-circular as well as 0-180$^o$ rotatable linear polarizations;

• The beamline optics adopts the scheme of PGM operating in collimated light, which allows convenient optimization of the $C_{ff}$ parameter for maximal flux or resolution. The horizontal focusing on the exit slit is performed directly by the CM through the cylinder FM. An ultimate $E/\Delta E$ above 33000 at 1 keV photon energy is achieved;

• The measures to increase the beamline transmission (including glancing incidence on the mirrors, minimal grating groove densities tightly matching the required resolution, appropriate choice of the blazed vs lamellar gratings and with optimized profile parameters, operation at flux-optimal $C_{ff}$ increasing with groove density) as well as the parameters of the ring optimized for the soft-X-ray region result in experimentally confirmed high photon flux up to $1 \times 10^{13}$ photons/s/0.01%BW at 1 keV photon energy;

• A beamline focalization method is introduced to coordinate motions of the CM and FM in one DOF and ensure fast focalization accompanied by maximal transmission. The beamline alignment was facilitated by the beam monitor allowing on-line evaluation of the horizontal beam position and profile at the exit slit. An overview of the EPICS-based beamline control system is given;

• Our ARPES endstation features a grazing light incidence angle of 20$^o$, horizontal axis of primary rotation, manipulator having 3 angular DOFs, and rotatable photoelectron analyzer. The RIXS endstation is equipped by the high-resolution spectrometer SAXES delivering $E/\Delta E \sim 11000$ at photon energy of 1 keV. It is mounted on a rotating platform to vary the scattering angle. Ellipsoidal refocusing optics demagnifies the vertical spot size below 4 μm, allowing slitless operation of the spectrometer. Due to low crossection of the soft-X-ray ARPES and RIXS the performance of the two endstations critically depends on the high flux delivered by the beamline.

## Acknowledgments

We thank Frisoe van der Veen, C. Quitmann and J. Mesot (Paul Scherrer Institute) and G. Ghiringhelli, C. Dallera and L. Braicovich (Politecnico di Milano) for their scientific and logistic support of the ADRESS project and promoting discussions. The SAXES spectrometer is developed jointly by Politecnico di Milano, Swiss Light Source and Ecole Polytechnique Fédéderale de Lausanne.

## References

[*] Email vladimir.strocov@psi.ch

[**] Also with Dipartimento di Fisica, Politecnico di Milano, Piazza Leonardo da Vinci 32, I-20133 Milano, Italy

[1]: T. Schmidt, A. Imhof, G. Ingold, B. Jakob and C. Vollenweider, *Proceedings of 9th International Conference on Synchrotron Radiation Instrumentation, Daegu* (2006) 400

[2] S. Sasaki *et al*, Nucl. Instr. and Meth. **A331** (1993) 763

[3] R. Carr, Nucl. Instr. and Meth. **A306** (1991) 391; S. Lidia and R. Carr, Nucl. Instr. and Meth. **A347** (1994) 77

[4] R. Follath and F. Senf, Nucl. Instr. and Meth. A **390** (1997) 388

[5] H. Petersen, C. Jung, C. Hellwig, W.B. Peatman and W. Gudat, Rev. Sci. Instr. **66** (1995) 1

[6] W.B. Peatman, *Gratings, Mirrors and Slits - Beamline Design for Soft-X-Ray Synchrotron Radiation Sources* (Gordon and Breach, 1997)

[7] H. Petersen, Optics Communications **40** (1982) 260

[8] F. Polack, B. Lagarde and M. Idir, AIP Conf. Proc. **879** (2007) 655

[9] U. Flechsig, L. Patthey and C. Quitmann, Nucl. Instr. and Meth. A **467-468** (2001) 479; C. Quitmann, U. Flechsig, L. Patthey, T. Schmidt, G. Ingold, M. Howells, M. Janousch and R. Abela, Surf. Sci. **480** (2001) 173

[10] J. Bahrdt, U. Flechsig, and F. Senf, Rev. Sci. Instr. **66** (1995) 2719

[11] J.B. West and H.A. Padmore, *Optical Engineering*, in *Handbook on Synchrotron Radiation* **2**, ed. by G.V. Marr (Elsevier, 1987)

[12] R. Follath, F. Senf and W. Gudat, J. Synchrotron Rad. **5** (1998) 769

[13] C.T. Chen, F. Sette, Rev. Sci. Instr. **60** (1989) 1616

[15] M. Kato, Y. Morishita, M. Oura, H. Yamaoka, Y. Tamenory, K. Okada, T. Matsudo, T. Gejo, I.H. Suzuki and N. Saito, in *Synchrotron Radiation Instrumentation: Ninth International Conference*, ed. by J.-Y. Choi and S. Rah (AIP, 2007)

[16] T. Tanaka, H. Shindo, C. Makochekanwa, M. Kitajima, H. Tanaka, A. De Fanis, Y. Tamenori, K. Okada, R. Feifel, S. Sorensen, E. Kukk and K. Ueda, Phys. Rev. A **72** (2005) 022507

[14] Available online at http://slsbl.web.psi.ch/beamlines/nitrogen-resolution.shtml

[17] *REFLEC, a program to calculate VUV/X-ray optical elements and synchrotron radiation beamline*, F. Schaefers, D. Abramsohn and M. Krumrey (BESSY, 2002). We thank BESSY for making the REFLEC code available to us. The code is based on the method described in M. Nevière, P. Vincent and D. Maystre, Appl. Optics **17** (1978) 843

[19] Y. Saitoh, H. Kimura, Y. Suzuki, T. Nakatani, T. Matsushita, T. Muro, T. Miyahara, M. Fujisawa, K. Soda, S. Ueda, H. Harada, M. Kotsugi, A. Sekiyama and S. Suga, Rev. Sci. Instrum. **71** (2000) 3254

[18] Available at www.ird-inc.com

[20] Product information available at http://www.fmb-oxford.com/product.php?product=53

[21] T.M. Mooney et al., Rev. Sci. Instrum. **67** (1996) 3369

[22] G. Jansa, R. Gajsek and M. Kobal, in *Proceedings of the 6th International Workshop on Personal Computers and Particle Accelerator Controls*, ed. by M. Bickley and P. Chevtsov (2007). Available at http://conferences.jlab.org/PCaPAC/PCaPAC2006_proceedings.pdf

[23] S. Staudenman; SLS Beamline Equipment Protection System, Technical note TM-93-01-01 (2001)

[24] A. Kotani and S. Shin, Rev. Mod. Phys. **73** (2001) 2003

[25] G. Ghiringhelli, A. Piazzalunga, C. Dallera, G. Trezzi, L. Braicovich, T. Schmitt, V.N. Strocov, R. Betemps, L. Patthey, X. Wang and M. Grioni, Rev. Sci. Instrum. **77** (2006) 113108

[26] V.N. Strocov, T. Schmitt and L. Patthey, PSI Technical Report No. SLS-SPC-TA-2008-309 (2008)

[27] G. Ghiringhelli, A. Piazzalunga, C. Dallera, T. Schmitt, V. N. Strocov, J. Schlappa, L. Patthey, X. Wang, H. Berger and M. Grioni, Phys. Rev. Lett. **102** (2009) 027401

[28] J. Schlappa, T. Schmitt, F. Vernay, V. N. Strocov, V. Ilakovac, B. Thielemann, H. M. Rønnow, S. Vanishri, A. Piazzalunga, X. Wang, L. Braicovich, G. Ghiringhelli, C. Marin, J. Mesot, B. Delley and L. Patthey, Phys. Rev. Lett. **103** (2009) 047401

**Tables**

Table 1. Parameters of the optical elements. The CM and PM are internally water cooled, the gratings side cooled, and other optical elements are not cooled. The CM was produced by ZEISS out of a substrate with cooling channels delivered by INSYNC. The PM, gratings and RMs are produced by ZEISS, and the FM by SESO.

|  | Shape | Grazing incidence angle (deg) | Meridional/ sagittal radii (mm) | Measured meridional/ sagittal *rms* slope errors (arcsec) | Substrate/ Coating |
|---|---|---|---|---|---|
| CM | Toroid | 1.5 | 573600/ 929 | 0.4/2.5 | Si/Pt |
| PM | Plane | - | inf | 0.052/0.094 | Si/Pt |
| Grating 800 l/mm | Plane | - | inf | 0.07/0.15 | Si/Au |
| Grating 2000 l/mm | Plane | - | inf | 0.066/0.08 | Si/Au |
| Grating 4200 l/mm | Plane | - | inf | 0.058/0.062 | Si/Au |
| FM | Cylinder | 1.0 | inf/332 | 0.29/1.4 | Si/Pt |
| RM1 | Toroid | 1.0 | 74000/ 24.1 | 0.58/0.78 | Fused silica/Pt |
| RM2 | Ellipsoid | 1.0 | Shape parameters: A=3425 mm, B=C=42.2 mm | 1.5/3.2 | Fused silica/Pt |

**Figure captions**

Fig. 1. The flux produced by UE44 within the central cone for a current of 400 mA in the ring for the linear horizontal and vertical (LH and LV) and circular (C) polarizations. The linear polarizations follow the same flux curve, but start from different energies.

Fig. 2. Optical layout of the beamline. The optical scheme is the collimated-light PGM.

Fig. 3. Theoretical $E/\Delta E$, including the spot size and diffraction contribution, as a function of photon energy. The three bunches of lines correspond to the three gratings (indicated on the right) and the individual lines within each bunch to variation of $C_{ff}$ by integer numbers (within the limits indicated on the left). The shaded bands display $E/\Delta E$ without the diffraction contribution, with their edges corresponding to the $C_{ff}$ limits of the bunches. The insert shows a typical spot at the exit slit.

Fig. 4. A few experimental X-ray absorption spectra of gases characteristic of energy resolution: (*a*) N $1s \rightarrow \pi^*$ resonance of $N_2$ (grating 800 l/mm); (*b*) $1s$ Rydberg series of Ne (800 l/mm); (*c*) O $1s \rightarrow 2\pi$ resonance of CO (4200 l/mm).

Fig. 5. Calculated reflectivity of blazed and lamellar gratings with $N$ = 800 l/mm and 2000 l/mm in the case of ideal $90^o$ apex angle (*solid lines*) and realistic obtuse one (*dashed*) as shown on the right. The profile parameters are optimized for the energy 930 eV and $C_{ff}$ = 2.25. The blazing has larger effect at smaller $N$.

Fig. 6. Optimization of the grating parameters for the 2000 l/mm grating: Transmission of the PM and grating pair $R_2$ over the energy interval 700 – 1200 eV, average $<R_2>$ (*top*) and relative variation $<\Delta R_2>/<R_2>$ (*bottom*). The crossing planes show the position of their maximum and minimum, respectively. The optimal values of $h$ and $c/d$ are chosen slightly shifted from the exact transmission maximum to improve flatness of the energy dependence.

Fig. 7. Theoretical flux at the sample as a function of photon energy for LH polarized light from the undulator. The three bunches of solid lines correspond to the three gratings (indicated on the right) and the bunch shown shaded to the blazed 800/mm grating operating in the $2^{nd}$ order. Variations of flux from the bottom to top edges of each bunch corresponds to variation of $C_{ff}$ in integer values within the limits indicated on the right. The flux is normalized to the bandwidth corresponding to the $E/\Delta E$ values typical of each grating (indicated on the right).

Fig. 8. Calculated energy dependences of the flux-optimal $C_{ff}$.

Fig. 9. Theoretical vertical (*top panel*) and horizontal (*bottom*) FWHM spot size at the sample as a function of the nominal demagnification $r/r'$ for refocusing with a typical toroidal and ellipsoidal mirror. Due to reduced aberrations the latter allows much smaller spot size when increasing $r/r'$.

Fig. 10. Image processing software to evaluate the beam profile at the exit slit: (*left panel*) Image with the region of interest containing the vertical line of energy dispersed light; (*right*) Horizontal beam profile and its Gaussian fit yielding the horizontal beam position and width.

Fig. 11. Vertical focusing scheme. The CM pitch $R_y^{CM}$, FM pitch $R_y^{FM}$ and FM transverse translation $z^{FM}$ combine in one focalization motion.

Fig. 12. Typical focalization curve as a resolution factor $F_R$ depending on the CFM parameter $z^{FM}$ (800 l/mm grating, $C_{ff} = 2.15$). Cubic polynomial fit of the measured points is shown by gray line. The insert shows the region in the O 1s absorption spectrum of CO used to define $F_R$.

Fig. 13. Scheme of the beamline control system.

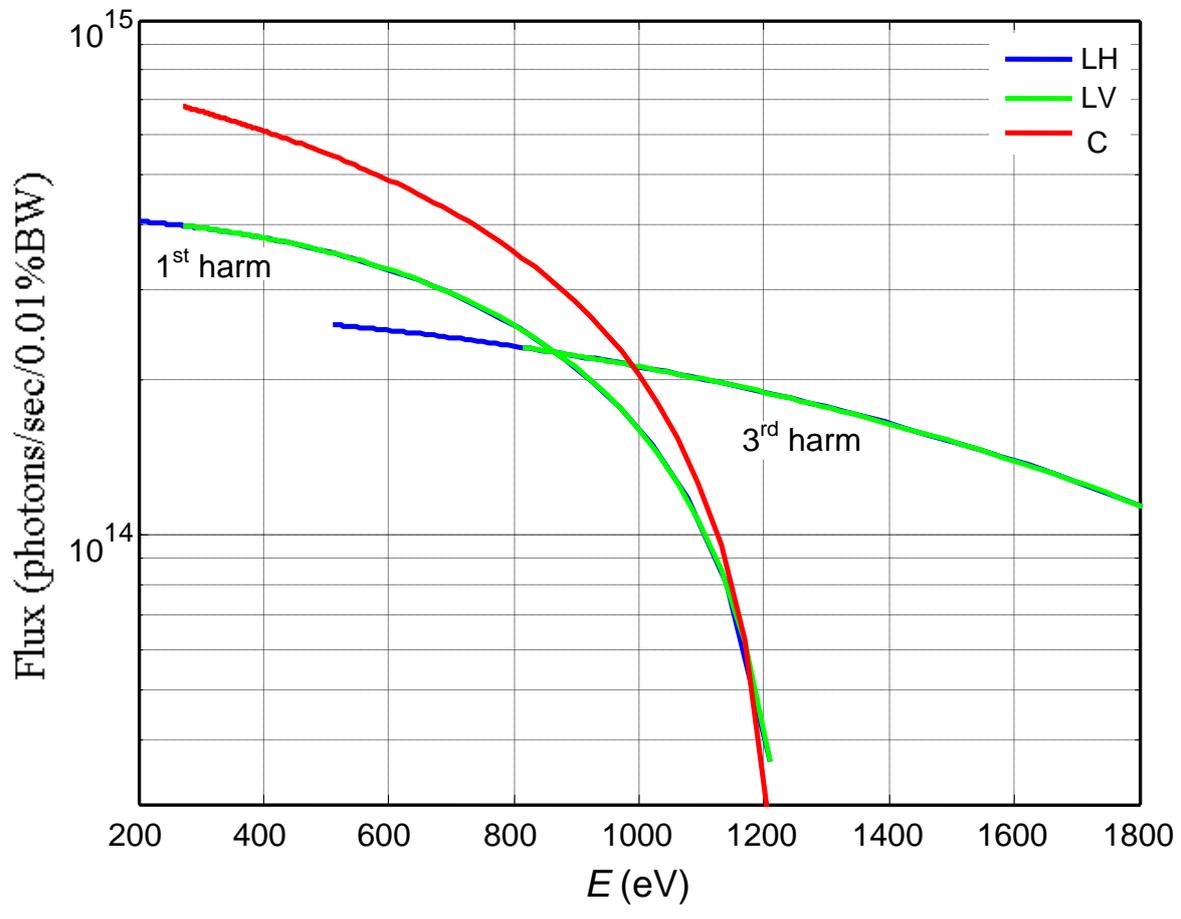

Fig. 1

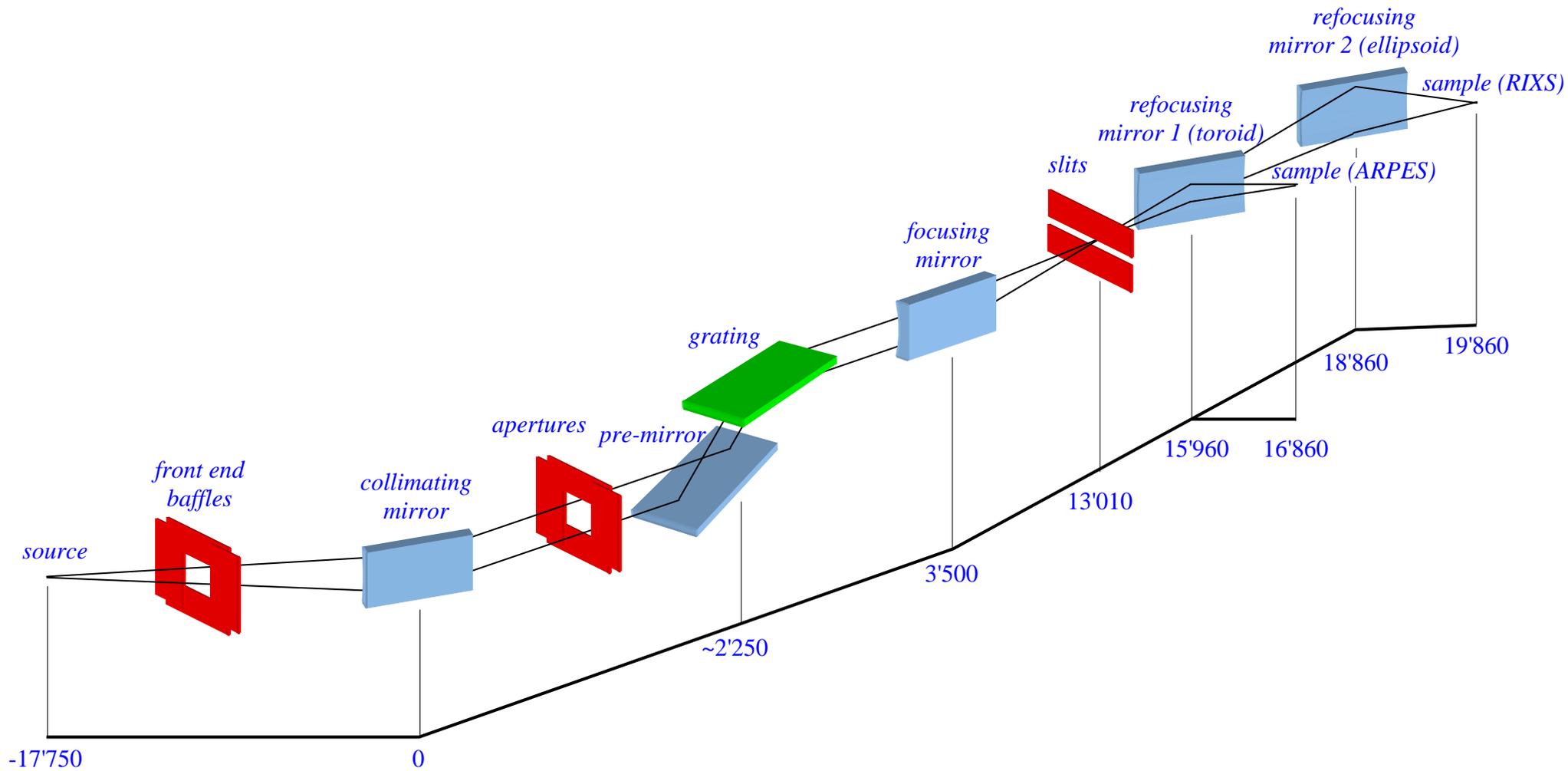

Fig. 2

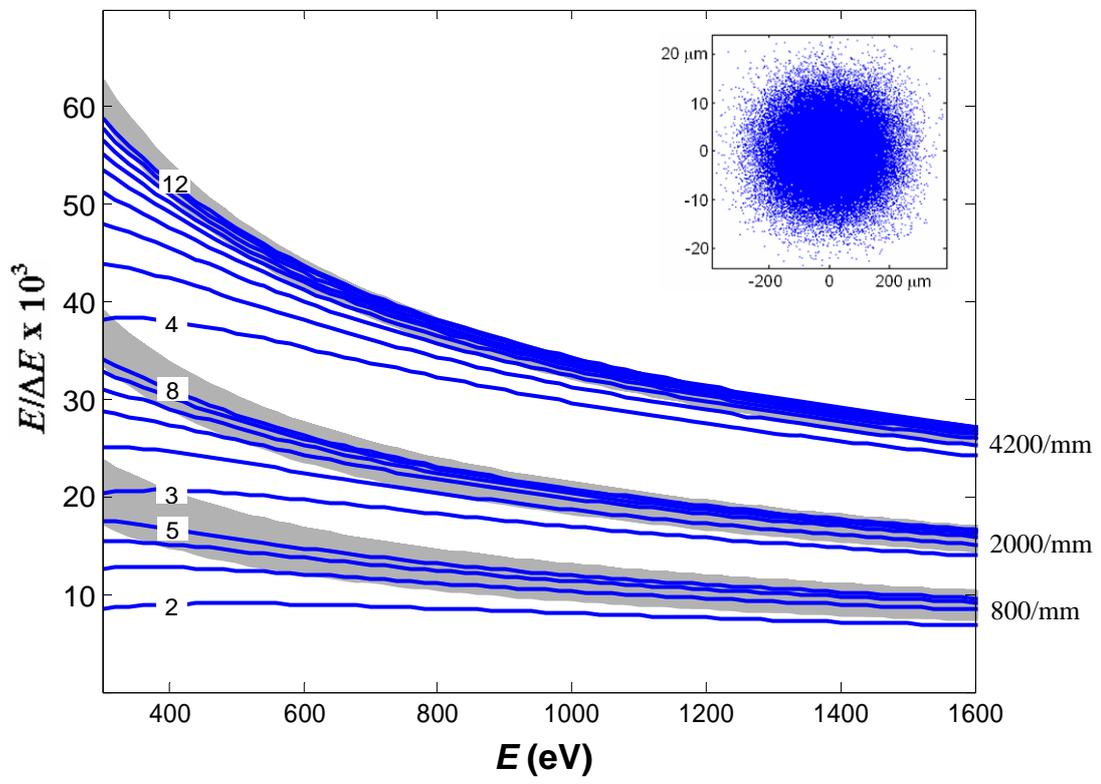

Fig. 3

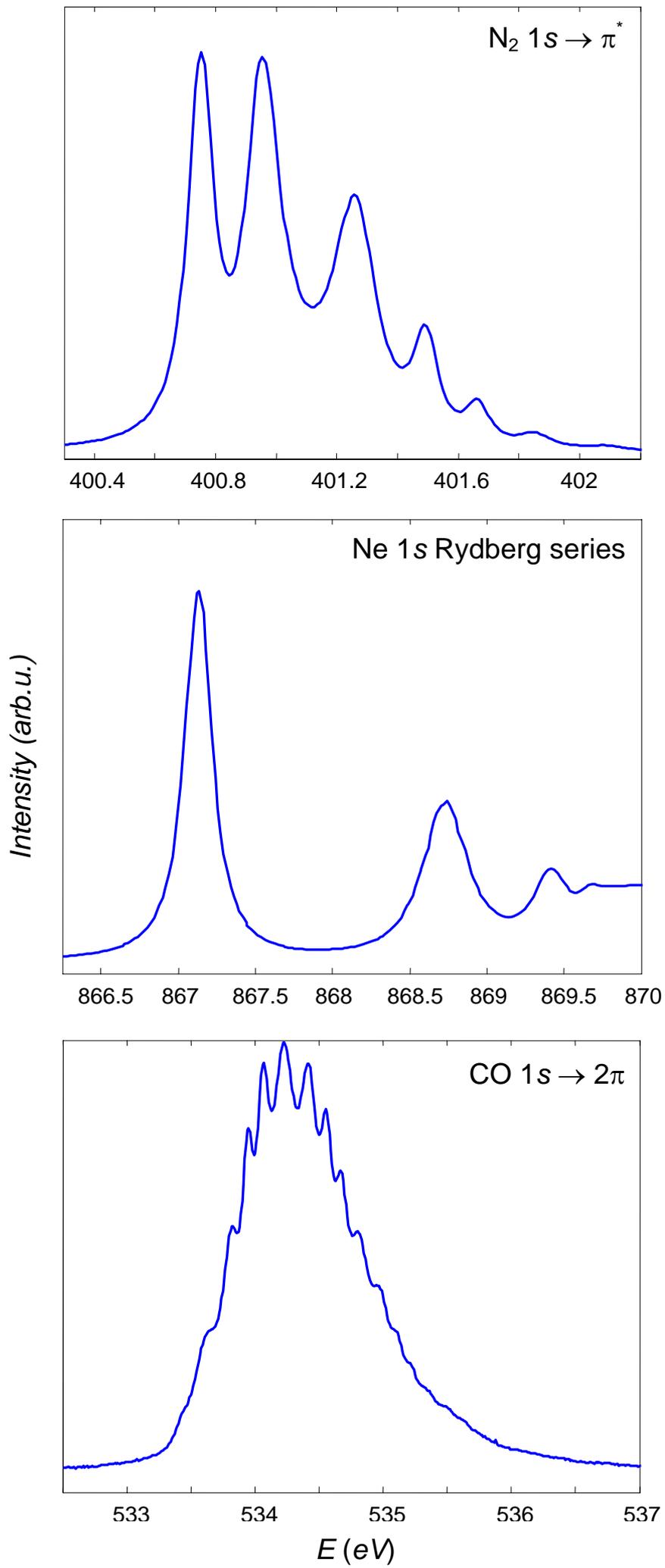

Fig. 4

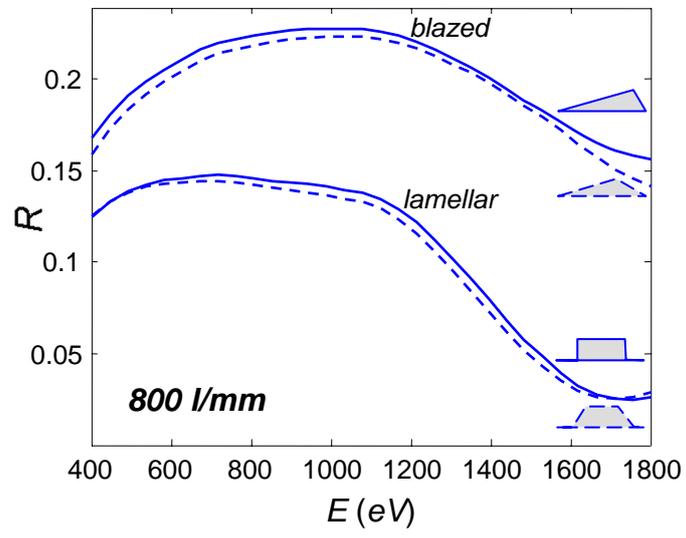
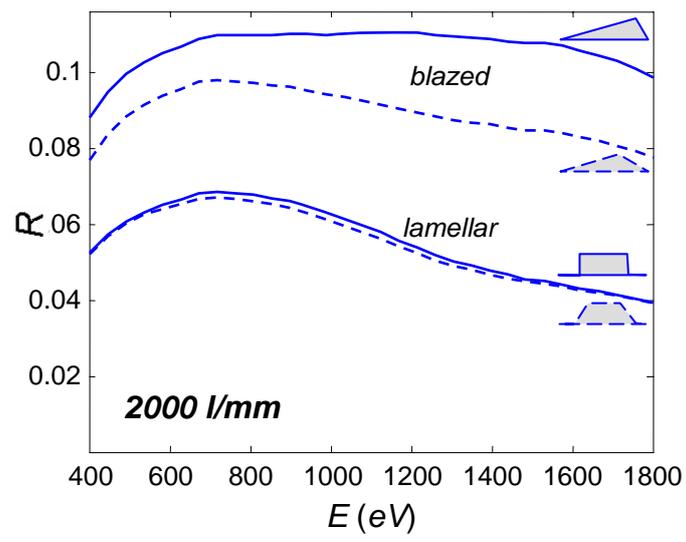

Fig. 5

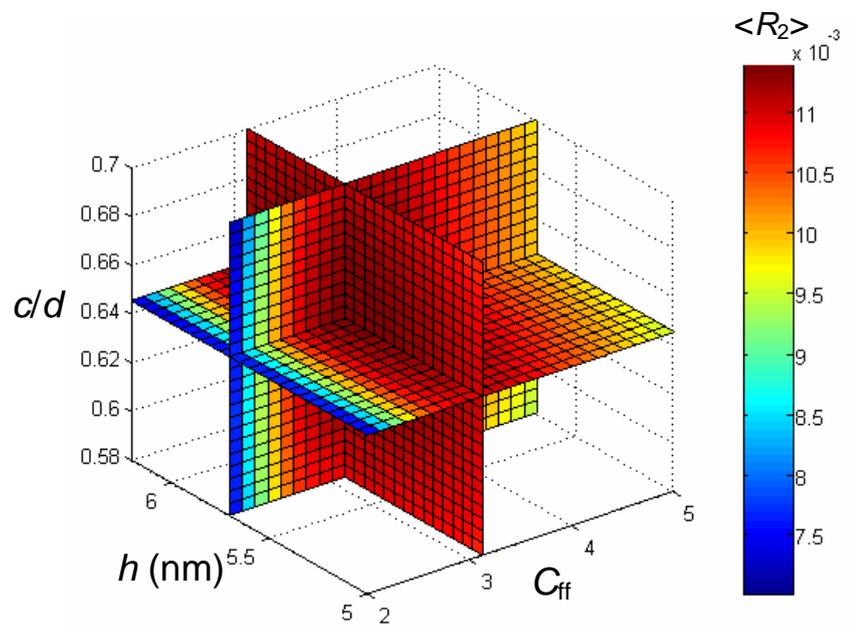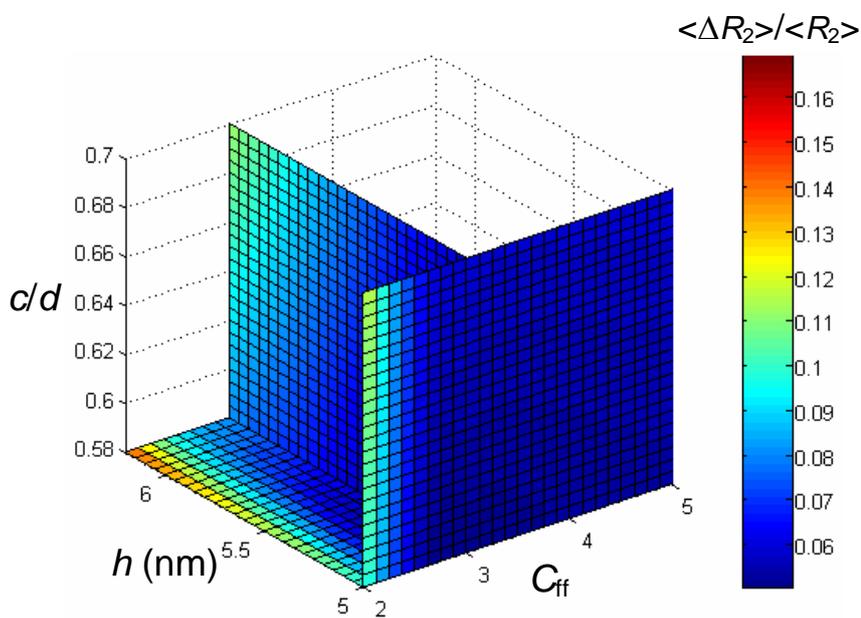

Fig. 6

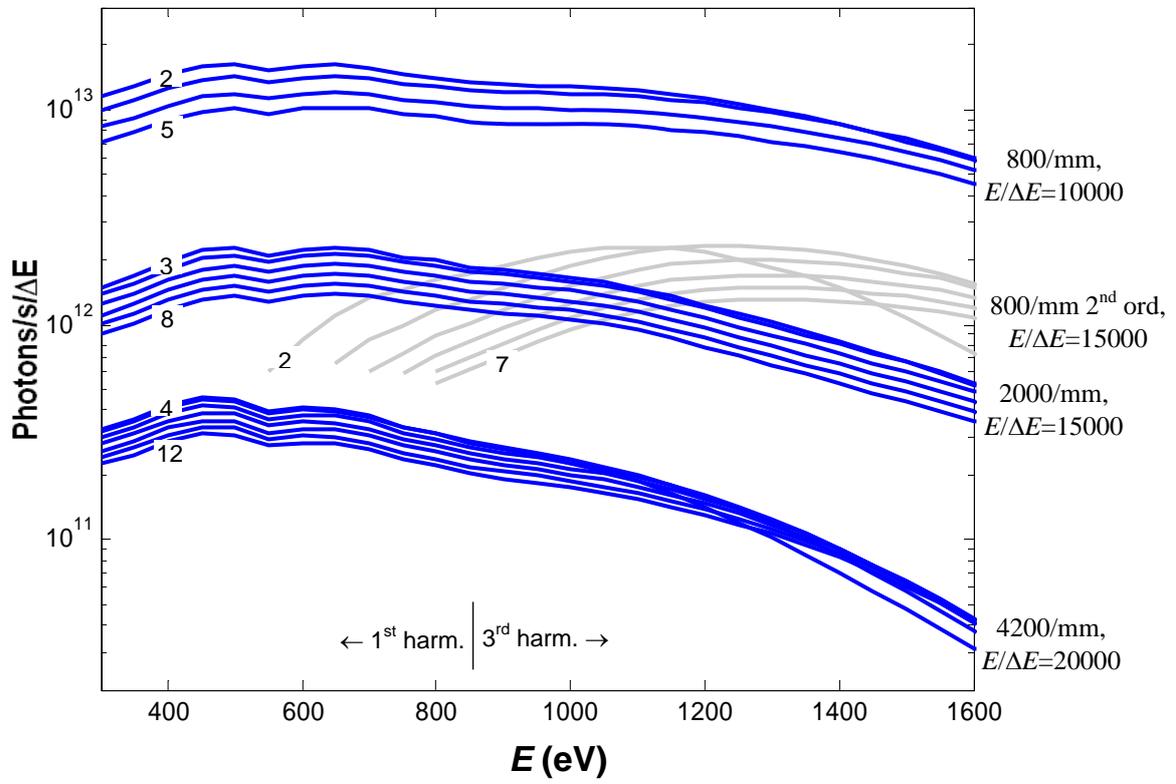

Fig. 7

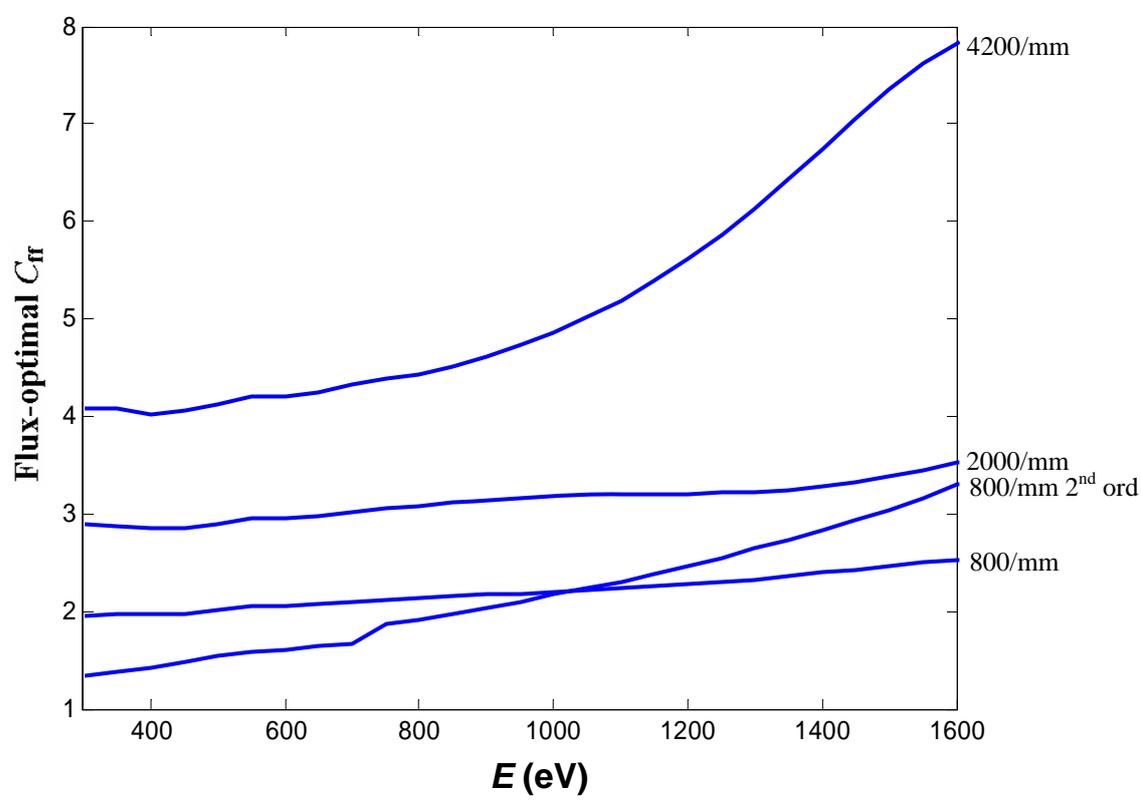

Fig. 8

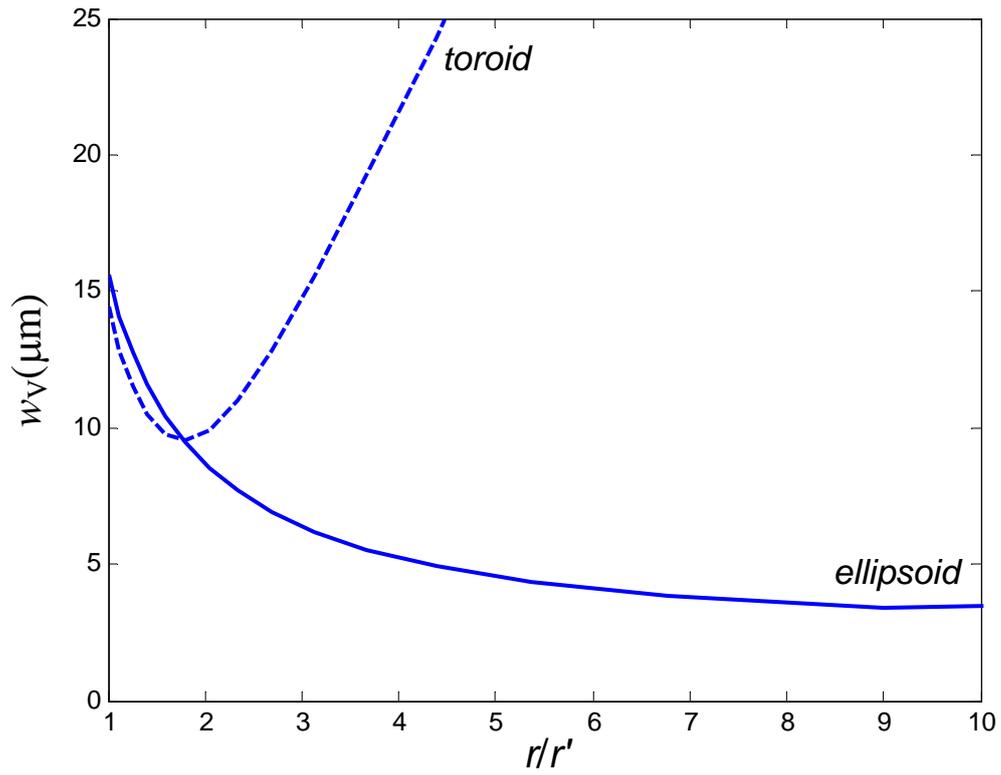

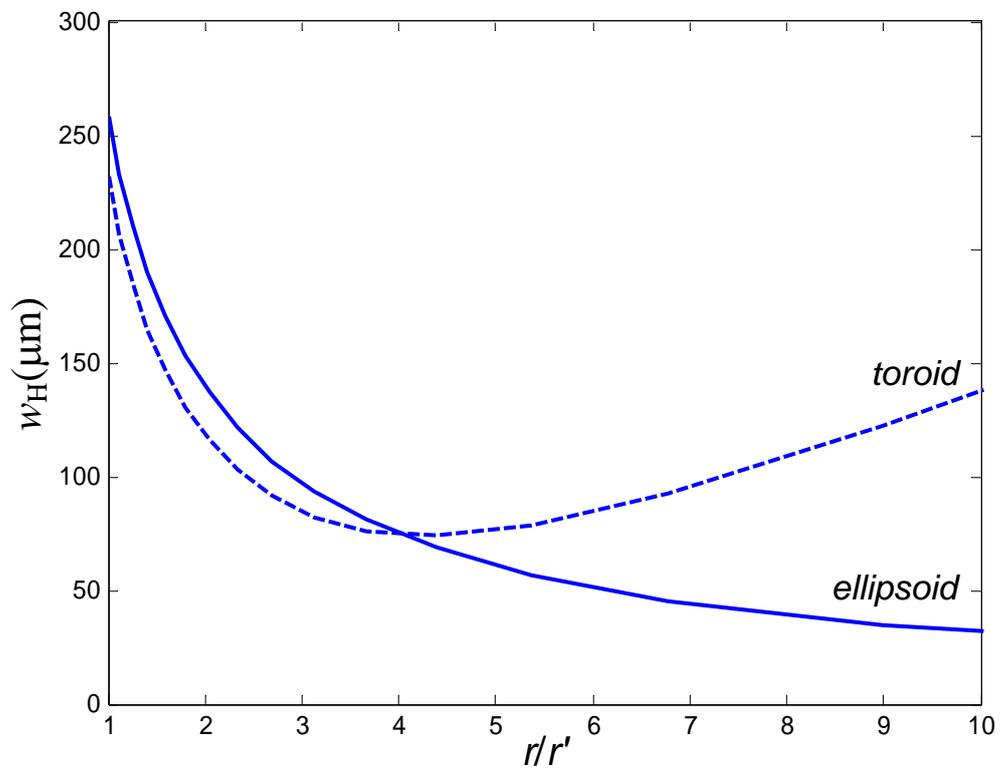

Fig. 9

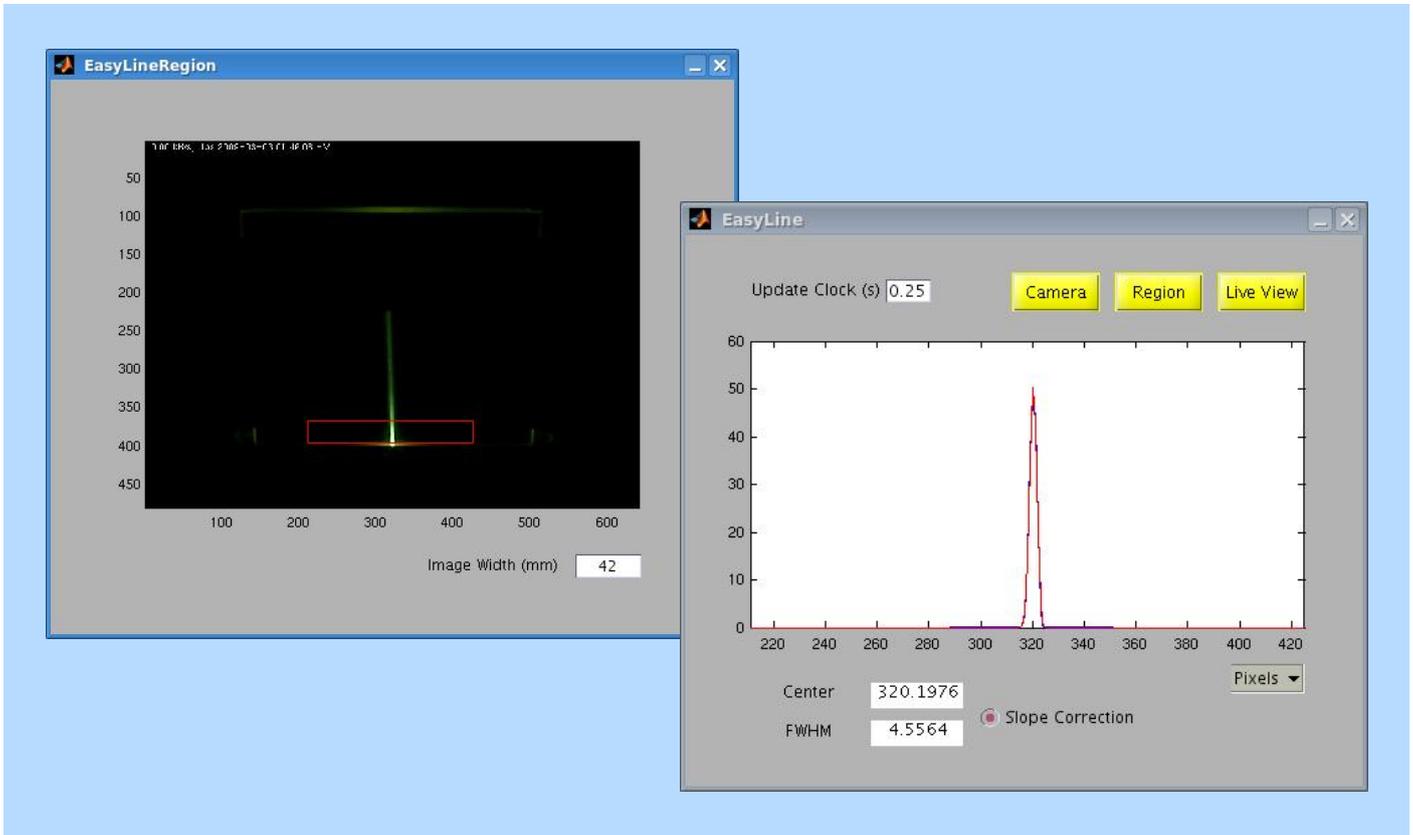

Fig. 10

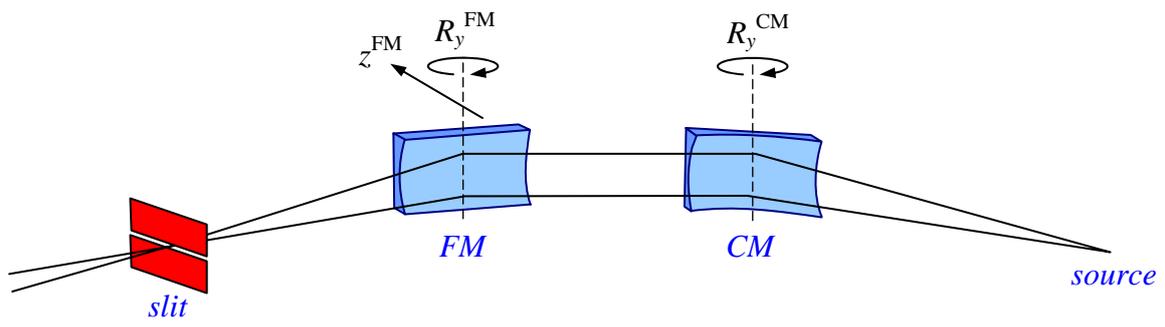

Fig. 11

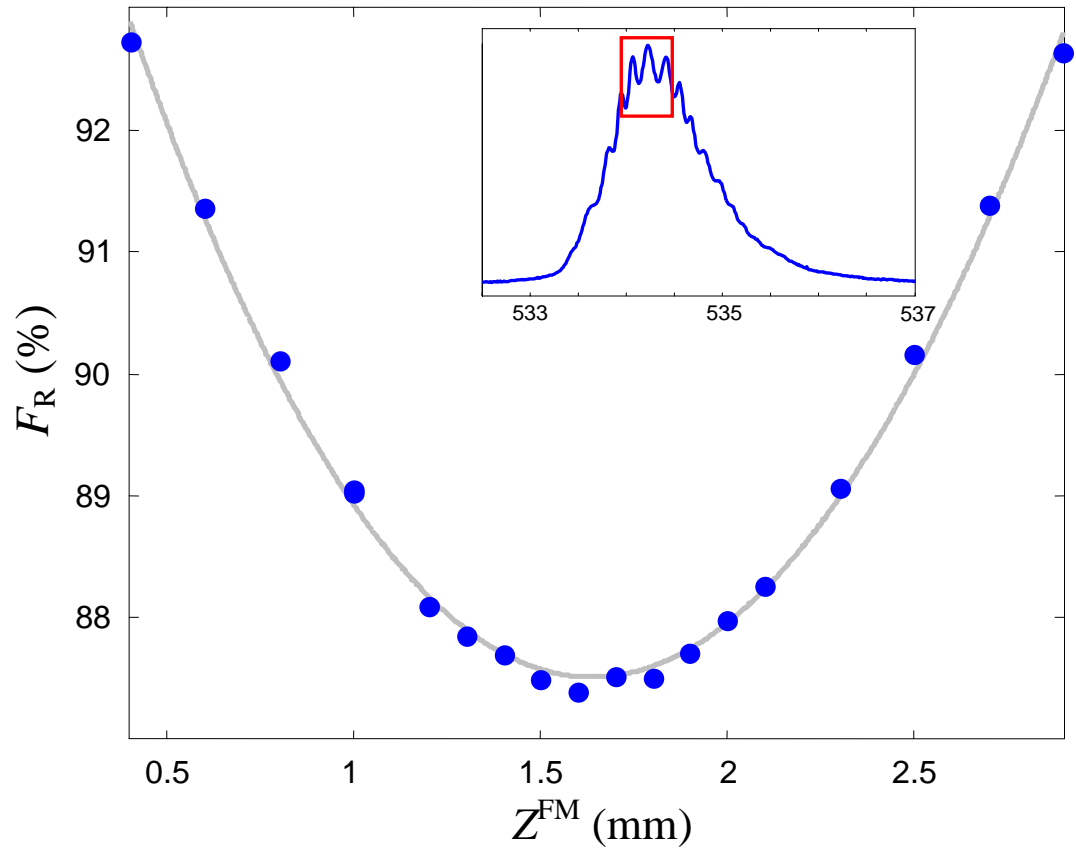

Fig. 12

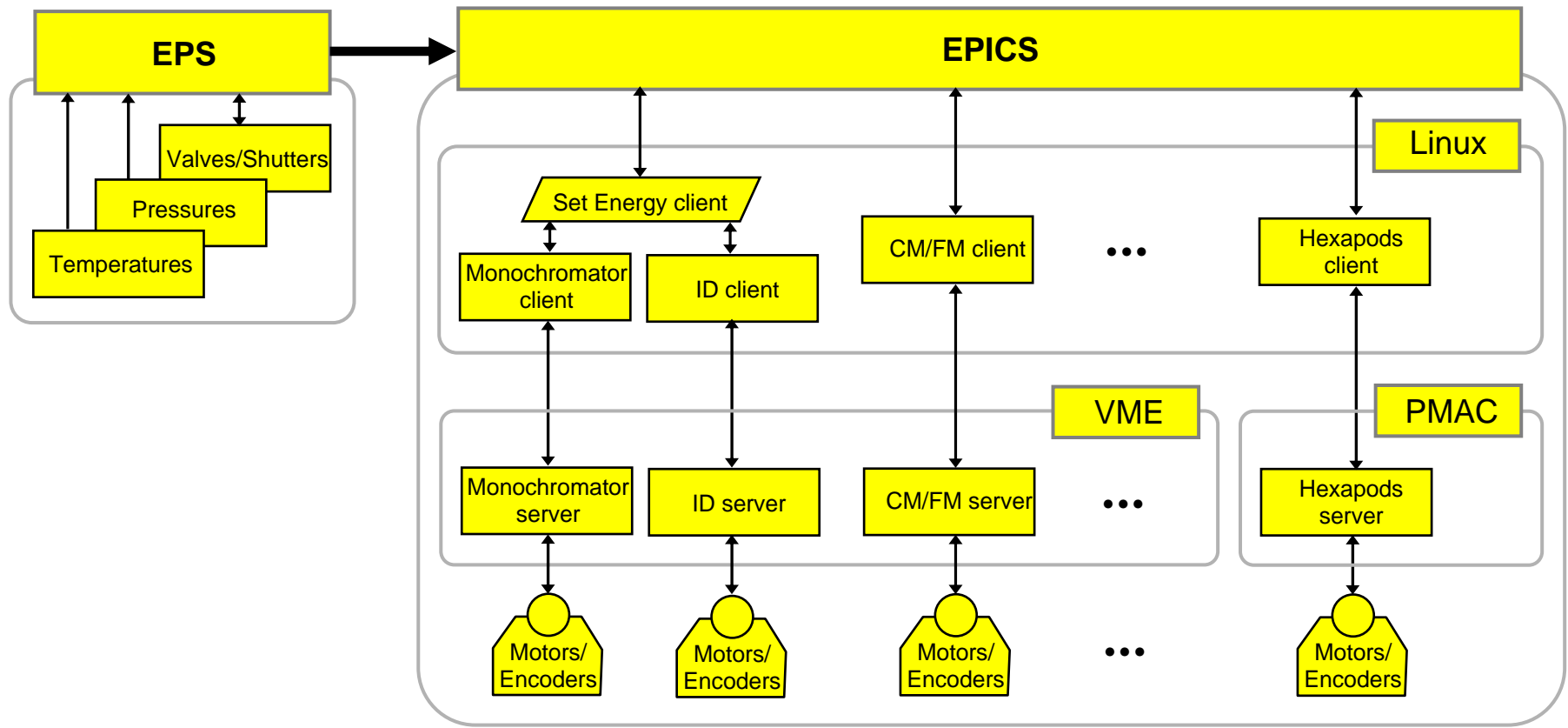

Fig. 13